\documentclass[a4paper,11pt]{article}
\usepackage{pos}
\usepackage{bbold}
\usepackage{cleveref}
\usepackage[backend=biber,sorting=none,maxnames=6, giveninits=true,url=false,]{biblatex}
\newcommand{\ijab}{\substack{
    i,j \\\alpha,\beta}}

\usepackage{placeins}

\usepackage[firstpageonly=true,angle=270,scale=0.2, fontsize=20pt,color=gray, pos={22cm,7cm}]{draftwatermark}
\SetWatermarkText{MS-TP-24-02}
\SetWatermarkScale{1}

\title{Optimized Distillation Profiles for Heavy-Light~Spectroscopy}
\author*[a]{Jan Neuendorf}
\author[a]{Giulia Egbring}
\author[a]{Jochen Heitger}
\author[b]{Roman Höllwieser}
\author[b]{Francesco Knechtli}
\author[b]{Tomasz Korzec}
\author[b]{ Juan Andrés Urrea-Niño}

\affiliation[a]{
Institut für Theoretische Physik, Universität Münster,
Wilhelm-Klemm-Street 9, 48149 Münster, Germany}

\affiliation[b]{Department of Physics, University of Wuppertal, Gaußstrasse 20, 42119 Germany}

\emailAdd{jan.neuendorf@uni-muenster.de}
\emailAdd{g\_egbr02@uni-muenster.de}
\emailAdd{heitger@uni-muenster.de}
\emailAdd{roman.hoellwieser@uni-wuppertal.de}
\emailAdd{knechtli@physik.uni-wuppertal.de}
\emailAdd{korzec@uni-wuppertal.de}
\emailAdd{urreanino@uni-wuppertal.de}

\abstract{It has been demonstrated that distillation profiles can be employed
to build optimized quarkonium interpolators for spectroscopy calculations
in lattice QCD. We test their usefulness for heavy-light systems on
(3+1)-flavor ensembles with mass-degenerate light and a charm quark in
the sea in preparation for a future $D\bar{D}$-scattering analysis.
The additional cost of light inversions naturally leads to the question
if knowledge of optimal profiles can be used to avoid superfluous
computations. We show such optimal profiles for different lattice sizes and pion masses and discuss general
trends.
Furthermore, we discuss the handling of momenta in this framework.}

\FullConference{%
  The 40th International Symposium on Lattice Field Theory (Lattice2023)\\
  July 31 to August 4, 2023\\
  Fermilab, Batavia, Illinois, USA
}

\begin{document}
\maketitle

\section{Introduction}
Distillation, as introduced in \cite{peardon_novel_2009}, has quickly established itself as an important tool for hadron spectroscopy. In its standard form however, it does not provide a degree of freedom that could be used to optimize the interpolators. This can be rectified with the introduction of distillation profiles~\cite{PhysRevD.106.034501}. 
Here we shall recapitulate and give a short summary of this approach. 
The distillation operator is usually written as 
\begin{equation}
S(t)=V(t)\,J\,V^\dagger(t),
\end{equation}
where the $V$ correspond to the eigenvectors of the lattice Laplacian, and a common choice is $J=\mathbb{1}$. The key idea of \cite{PhysRevD.106.034501} is to use the diagonal matrix $J$
\begin{equation}
\label{dist_operator}
J_{\ijab}(t)=\delta_{ij}\,\delta_{\alpha\beta} \,{g(\lambda_i(t))}
\end{equation}
as an additional degree of freedom to be exploited in a variational formulation, 
where the entries of $J$ are a function of the Laplacian eigenvalue $\lambda$.
When calculating correlators
\begin{equation}
\label{correlator}
C_{\mathrm{2pt}}(t)=-\langle\mathrm{tr}\left[ \Phi_2(t)\,
			\tau_{q_a}(t,0)\,
			\bar{\Phi}_1(0)\,
			\tau_{q_b}(0,t)
			\right]\rangle_{\mathrm{gauge}},
\end{equation}
the \textit{perambulator} is introduced, as it was in \cite{peardon_novel_2009}, 
\begin{equation}
    \tau(t_1,t_2)=V^\dagger(t_1)\,D^{-1}\,V(t_2),
\end{equation}
and everything else is included in the \textit{elemental} 
\begin{equation}
\Phi_{\ijab}(t)=V_i^\dagger(t)\,\Gamma_{\alpha,\beta}(t)\,{g^\ast(\lambda_i(t))\,g(\lambda_j(t))}\,V_j(t).
\end{equation}
The important thing is that the perambulators are not changed by the introduction of $J\neq \mathbb{1}$. Therefore, none of the inversions of the Dirac operator have to be redone when $J$ is changed. In \cite{PhysRevD.106.034501} the functions $g(\lambda_i(t))$ are called quark profiles and it is demonstrated that a Gaussian with its peak at zero is a beneficial choice. This can be understood as follows. When using standard distillation, the Laplacian eigenmodes above a certain index $N_V$ are cut off and all lower eigenmodes are used equally. By introducing $J\neq \mathbb{1}$ the higher eigenmodes are still cut off, but even below threshold the entries approaching $N_V$ from below are suppressed. The extent of this suppression is determined by the width of the Gaussian profile, which is parameterized by $\sigma$
\begin{equation}
g_n(\lambda)=\exp(-\frac{\lambda^2}{\sigma_n^2}).
\end{equation}
In this equation $\sigma$ appears with an index, hinting at the fact that one can use a basis of multiple Gaussian profiles and use the GEVP \cite{LUSCHER1990222,blossier_efficient_2008} to optimize the interpolator. 
The result can be interpreted as an optimized linear combination of the Gaussian profiles called the optimal profile. 
In \cite{PhysRevD.106.034501} this method was successfully applied for a setup with unphysical degenerate heavy quarks. 
The purpose of this work is to demonstrate that this approach can also be used for the heavy-light system and to take a first look at the qualitative differences between profiles of heavy-light systems and charmonium. 
One reason why an exploration of heterogeneous mesons with this method is of interest is that the quark profile is not directly accessible, as it always appears in pairs in \cref{correlator}. In the case of charmonium, it is natural to assume that a symmetry relation between the two profiles of identical quarks exists, which is no longer the case for heavy-light systems.
In addition to this, we explore the application of distillation profiles at non-zero momenta, using both lattice momentum and partially twisted periodic boundary conditions \cite{aharonov_significance_1959,dedivitiis_discretization_2004}. 

\section{Ensembles and Computational Details}
Two very different ensembles are used in this work, both of which use the improved Wilson action \cite{sheikholeslami_improved_1985}. Their most important parameters are given in \cref{ensemble_tab}. 
\begin{table}
    \centering
   \begin{tabular}{l|c|c|c|c|c|c}
				Name&$N_f$&Temporal boundaries&$a$[fm]&$L^3\times T$&$N_V$&$m_\pi$[GeV]\\\hline
				A11&3+1&open&0.054&$32^3\times 96$&200&$\approx$ 1 or 0.42 \\
				D5&2&periodic&0.0653&$24^3\times48$&200&0.439
			\end{tabular}
   \caption{Parameters of the two ensembles used in this work. \label{ensemble_tab}}
\end{table}
The first ensemble (A11) is described in \cite{Hllwieser2020}. It is at the $SU(3)_{\rm flavor}$ symmetric point. 
There are two sub-variants of this ensemble which differ only in their pion mass. Here, only the lighter pion mass is used.
The second ensemble (D5) is a two-flavor ensemble provided by the the CLS (Coordinated Lattice Simulations) team and employed in earlier studies e.g. \cite{heitger_charm_2014,fritzsch_strange_2012,FINKENRATH2013441}. Its charm quark mass is chosen such that $m_{D_s} = m_{D_s,\text{phys}}$ \cite{heitger_charm_2014} and $\kappa_s$ is tuned to keep $\frac{m_K}{f_K}$ and $\frac{m_\pi}{f_K}$ physical \cite{fritzsch_strange_2012}. This ensemble was included to test the use of periodic boundary conditions in combination with distillation profiles as this is of interest to us for a project described in \cite{Neuendorf:2022WW}.
The computational setup used to obtain the perambulators is the same as in \cite{PhysRevD.106.034501} and $N_V=200$ is used for both ensembles since their 3D volumes in physical units are approximately equal.
The elementals and correlators are computed using a custom implementation. 
The package \texttt{pyerrors} \cite{JOSWIG2023108750} is used for its implementation of the $\Gamma$-Method for error estimation \cite{WOLFF2004143} and for solving the generalized eigenvalue problem.
For this exploratory project, only a part of the available statistics is used. 

\section{A First Look at the Profiles}
\FloatBarrier
\begin{figure}
    \centering
    \includegraphics[width=0.9\textwidth]{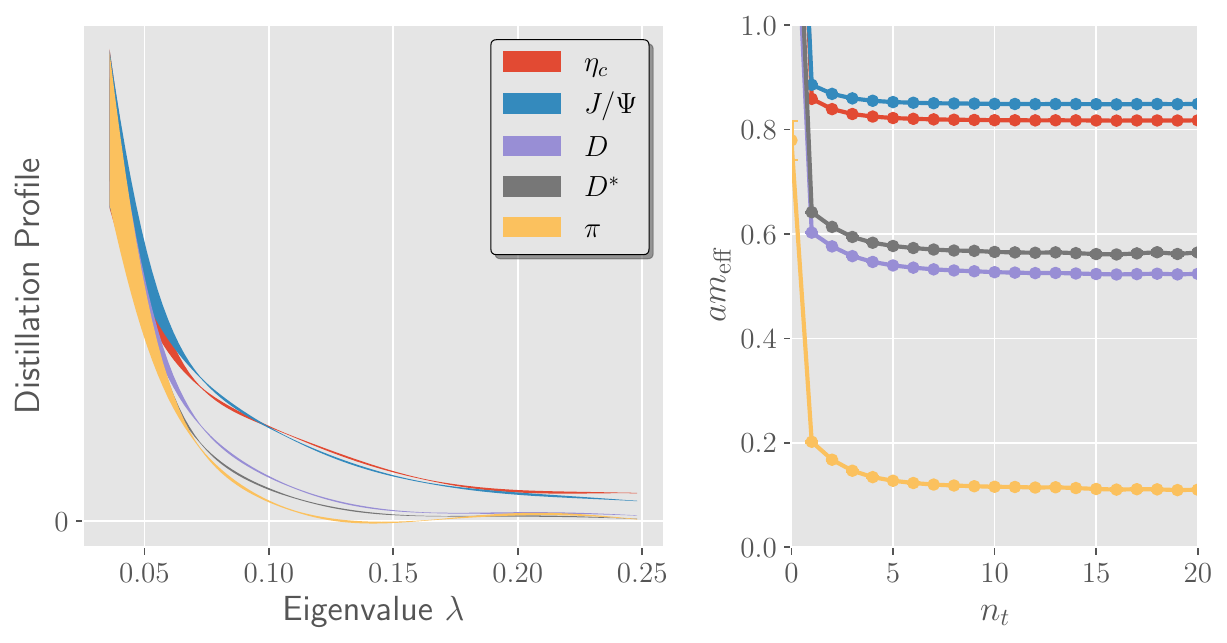}
    \caption{Optimized profiles \textit{(left)} and effective masses \textit{(right)} for different particles on the ensemble A11.}
    \label{fig_profiles_overview}
\end{figure}
\begin{figure}
    \centering
    \includegraphics[width=0.9\textwidth]{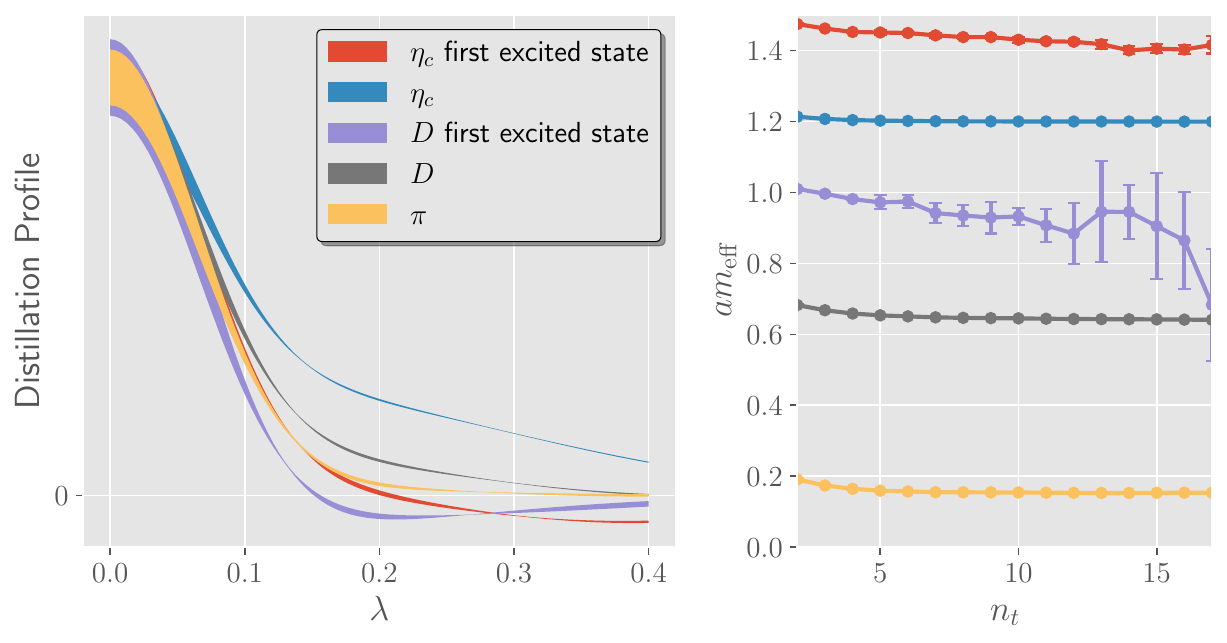}
    \caption{Optimized profiles \textit{(left)} and effective masses \textit{(right)} on CLS ensemble D5.}
    \label{fig_d5_profiles}
\end{figure}
 In \cref{fig_profiles_overview,fig_d5_profiles} optimal profiles and their resulting effective masses are shown for an assortment of particle channels. A few remarks are necessary regarding the portrayal and interpretation of the optimal profiles. 
 As the optimal profile is a visualization of a solution to the GEVP, it comes with an arbitrary amplitude and sign. Here, all profiles are fixed to start at the same positive value. The errors of the profiles derive from the errors of the eigenvectors. They include only the correlated error propagation from the operations used to solve the GEVP. %
 Since the profiles are divided by their starting value, the error of the ratio at that point would naturally be zero. We choose to re-scale the value and error of the profile to improve readability of the error bands.

 The plots of the profiles run from lower eigenmodes on the left to higher eigenmodes on the right. 
 In the language of smeared sources, the left side corresponds to contributions from smoother fields while the right side corresponds to more localized sources. Excited states often show nodes in their profiles. This was already observed in \cite{PhysRevD.106.034501}. In both \cref{fig_profiles_overview} and \cref{fig_d5_profiles} a clear trend can be seen. 
 The profiles of charmonium are wider than the ones of the $D$-meson, which themselves are wider than the ones of the pion. 
 Lighter particles seem to have narrower profiles and therefore a better overlap with more smeared sources. 
 
 \begin{figure}
    \centering
    \includegraphics[width=0.3\textwidth]{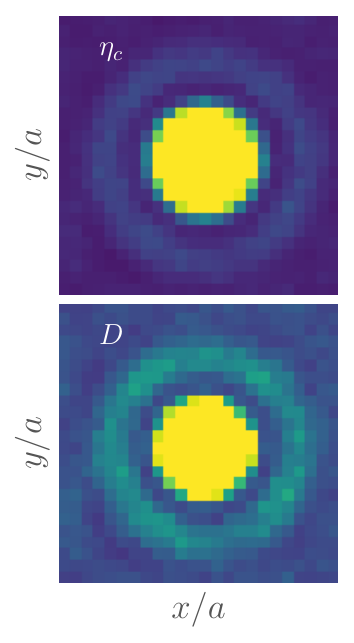}
    \includegraphics[width=0.6\textwidth]{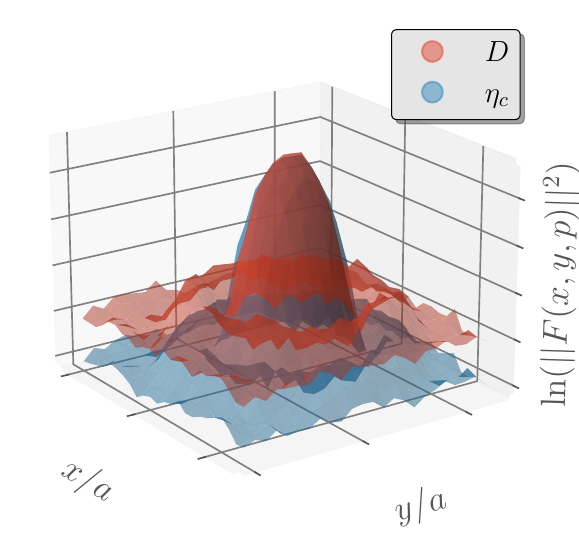}
    \caption{Visualization of the smeared quark fields from the optimal profiles on D5. A point source is introduced at the center of the lattice. Time and gauge configurations are averaged over. One spatial axis is sliced at the height of the point source.
    \textit{Left}: colormap of the normalized field. Bright colors are higher. High values are clamped at 0.03 .  \textit{Right}: logarithmic 3D plot using the same data as the left plots. }
    \label{fig_d5_profile_vis}
\end{figure}

From the optimal profiles one can work backwards to find the corresponding optimal spatial sources. 
This is done along the lines of \cite{PhysRevD.106.034501} by reconstructing the distillation operator (\cref{dist_operator}) and applying it to a point-source. The resulting field spans the volume of the lattice and still has Dirac and color indices. By averaging over color, applying $\Psi\to \rm{tr}(\gamma_5 \Psi)$ and taking the absolute value one obtains a real scalar field that can be plotted. This is done in \cref{fig_d5_profile_vis}. The result seems to confirm the observations from \cref{fig_profiles_overview,fig_d5_profiles} as the less localized background appears more significant for the heavy-light system. There are, however, two things to note. The plot is logarithmic and the central peak contains most of the field, and the plot shows concentric rings. Those rings are observed in \cite{PhysRevD.106.034501} as well but only for excited states.

\FloatBarrier
\section{Restricting $N_V^{\mathrm{light}}$}
\begin{figure}
    \centering
    \includegraphics[width=0.6\textwidth]{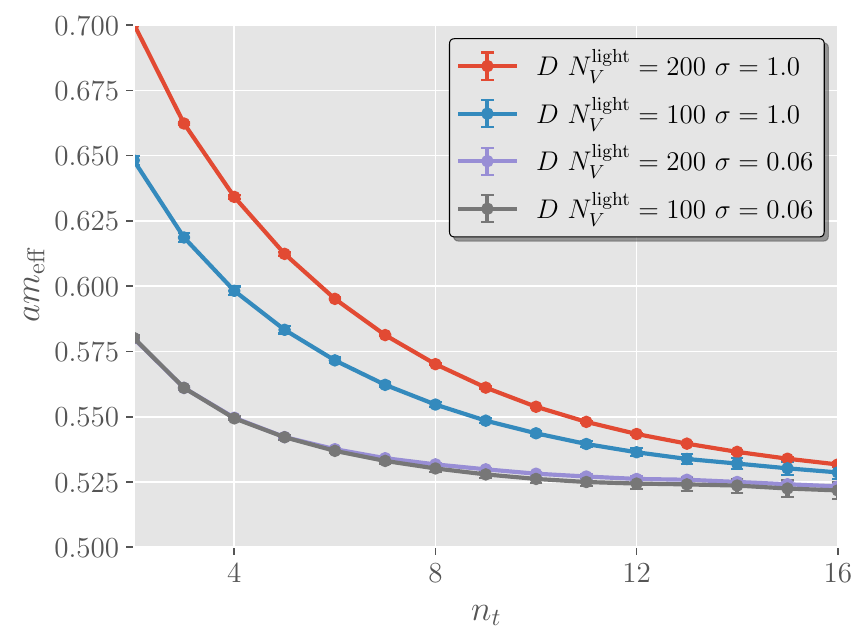}
    \caption{Effective $D$-meson masses for different values of $N_V^{\rm light}$.}
    \label{fig_nv_light}
\end{figure}

\FloatBarrier
The observation that the profiles of mesons containing lighter quarks are narrower might be a useful one. 
The number of inversions needed per configuration is given by $4N_V T$, where $T$ might be smaller than the temporal extent of the lattice in the case of open boundary conditions. Light inversions are costly and a natural approach to reduce this cost is to restrict $N_V$. The fact that higher eigenmodes are found to contribute very little to the optimal interpolator justifies this restriction. While all 200 inversion were performed, the restriction can be modeled by setting the perambulator to zero wherever one of its distillation indices exceeds a certain cutoff. This is only done for the light perambulator. Trivially, the change of the light perambulator does not effect charmonium. For the heavy-light system the results are shown in \cref{fig_nv_light}. They are almost compatible for Gaussian profiles with a small width and agree less as the width gets larger. This is to be expected, as only wider profiles are sensitive to larger eigenmodes. For the data shown here, a cut to $N_V^{\mathrm{light}}=100$ appears justified, cutting the cost of inversions almost in half. It can be shown that for elementals diagonal in eigenvector space only $\mathrm{min}\left\{N_V^{\rm light},N_V^{\rm charm}\right\}$ eigenvectors contribute to the correlators of D-mesons.

\FloatBarrier
\section{Momenta}
Lattice momenta can be introduced by using the general form of the elemental \cite{peardon_novel_2009,bruno_isospin_2023}
\begin{equation}
			\Phi_{i,j}(\vec{p})=\sum\limits_{\vec{x}}
			V_i^\dagger(\vec{x})\,e^{-i\vec{p}\cdot\vec{x}}\,g^\ast(\lambda_i)\,g(\lambda_j)\,\Gamma \,V_j(\vec{x}),
\end{equation}
suppressing Dirac indices.
This is appealing from a computational standpoint for two reasons. First, the perambulators are unaffected and no repeated inversions are required. Second, the elemental can be further split into a part depending on the profile and the gamma-structure and a part encoding the momentum, while the latter can be reused for all computations at the same lattice momentum:
\begin{equation}
\Phi_{i,j}(\vec{p})=\Gamma g(\lambda_i)g(\lambda_j)\sum\limits_{\vec{x}}
V_i^\dagger(\vec{x})\,e^{-i\vec{p}\cdot\vec{x}} \,V_j(\vec{x}).
\end{equation}
Note that $\Gamma$ is a $4\times4$ Dirac matrix that appears at every entry of the $N_V\times N_V$ elemental and that this factorization only holds for elementals without covariant derivatives.  
\begin{figure}
    \centering
    \includegraphics[width=0.9\textwidth]{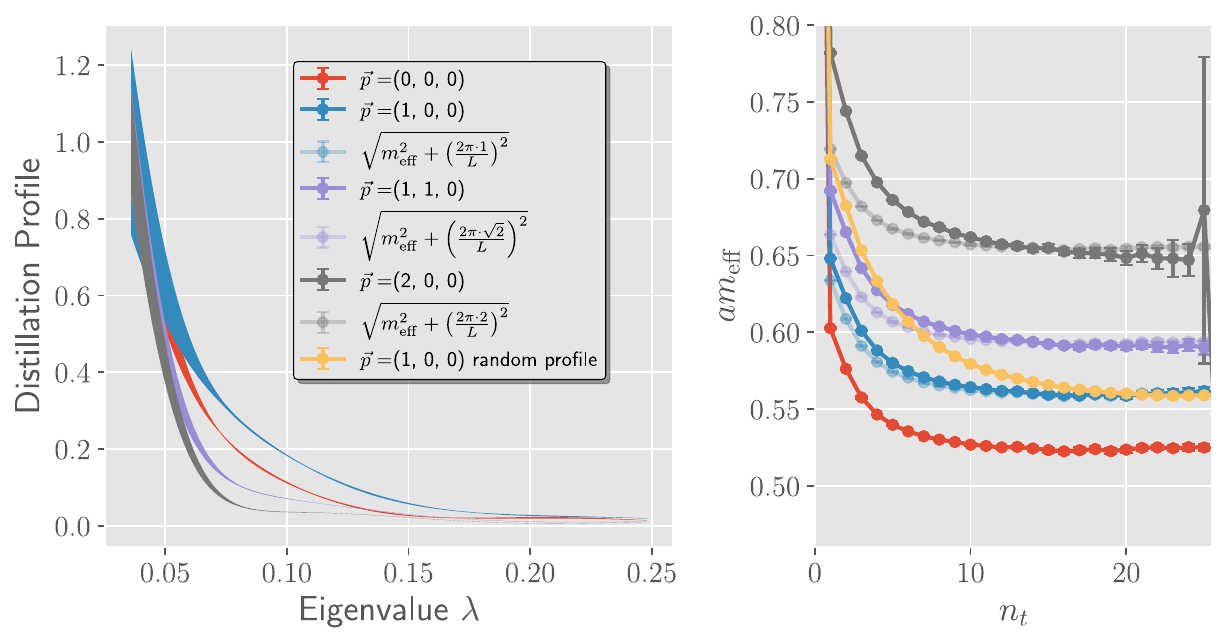}
    \caption{Optimized profiles \textit{(left)} and effective energies \textit{(right)} for different momenta on A11.}
    \label{fig_a1_momenta}
\end{figure}%
Results at non-zero momentum can be seen in \cref{fig_a1_momenta}. On the right side of the plots in transparent colors we show the rest-energy scaled by the continuum dispersion relation. Apart from a slower convergence to a plateau, this matches the energies at higher momenta relatively well. We also show a random profile at the first lattice momentum where all Gaussian profiles contribute equally, regardless of the GEVP. The result converges slower than the optimized profile, which can be seen as evidence that the optimization procedure still works well at non-zero momentum. Higher momenta were computed but are not shown as they become much noisier. 

\begin{figure}
    \centering
    \includegraphics[width=0.9\textwidth]{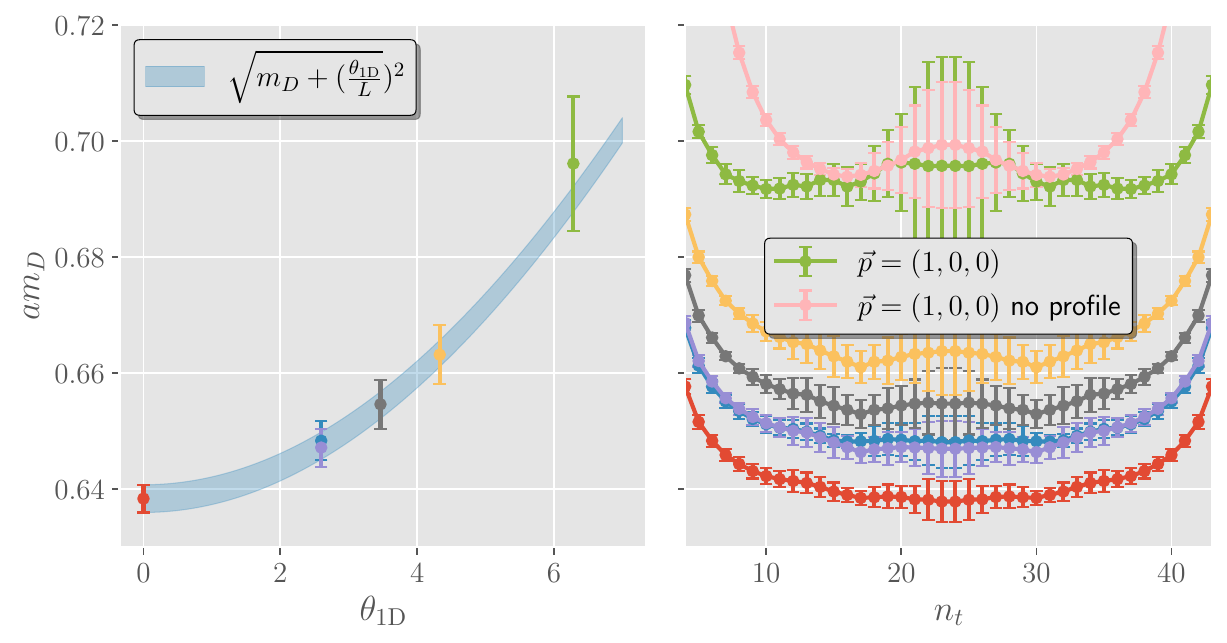}
    \caption{Dispersion \textit{(left)} and effective energies \textit{(right)} on D5 at different momenta, induced by lattice-momenta and isotropic partially twisted periodic boundary conditions. A profile with $\sigma= 0.1$ is used. For the second point in the left plot, $\theta=\pm 2.6$ is included for test purposes. The twisting angle is 2$\pi$ for the first lattice momentum and the continuum dispersion relation is displayed for comparison.}
    \label{fig_d5_momenta}
\end{figure}

To apply partially twisted periodic boundary conditions in this context one needs to recompute the perambulator with the desired boundary conditions applied. Here, the boundary conditions are only imposed on the charm quark. Then, one only needs to use this perambulator in place of the non-twisted one in the rest of the calculation. 
This is done for the ensemble D5. In \cref{fig_d5_momenta} we see that the momenta match the continuum dispersion relation. A twisting angle of $2\pi$ would correspond to the first lattice momentum, which is also included. Furthermore, a comparison is shown between the first lattice momentum with a Gaussian profile and standard distillation. One of the points is computed using both a positive and a negative twisting angle to show that they agree on the isotropic lattice.

\clearpage
\section{Summary}

The present investigation reveals the efficacy of distillation profiles across a range of scenarios, notably demonstrating their merit for heavy-light systems and at non-zero lattice momentum which has not been addressed so far. 
We find that the profiles tend to become narrow when lighter quarks are included. This leads to the idea of further restricting the number of light eigenmodes to save computation cost. The tests reported here were performed on two very different ensembles, one at the $SU(3)$ symmetric point with a dynamic charm quark and one with two degenerate light quarks and periodic boundary conditions. 

\section*{Acknowledgements}
The authors gratefully acknowledge the Gauss Centre for Supercomputing e.V. \\(\url{www.gauss-centre.eu}) for funding this project by providing computing time on the GCS Supercomputer SuperMUC-NG at Leibniz Supercomputing Centre (\url{www.lrz.de}) under GCS/LS project ID \textit{pn29se} as well as computing time and storage on the GCS Supercomputer JUWELS at Jülich Supercomputing Centre (JSC) under GCS/NIC project ID \textit{HWU35}. The authors also gratefully acknowledge the scientific support and HPC resources provided by the Erlangen National High Performance Computing Center (NHR@FAU) of the Friedrich-Alexander-Universität Erlangen-Nürnberg (FAU) under the NHR project \textit{k103bf}. This work is supported by the programme “Netzwerke 2021”, an initiative of the Ministry of Culture and Science of the State of Northrhine Westphalia, in the NRW-FAIR network,
funding code NW21-024-A and by the German Research Foundation (DFG) research unit FOR5269 “Future methods for studying confined gluons in QCD”.
\\
Additionally, this work is supported by the DFG through
the Research Training Group “GRK 2149: Strong and Weak Interactions – from Hadrons to Dark
Matter” (J.N. and J.H.).
Parts of the calculations for this publication were performed on the HPC cluster PALMA II of the University of Münster, subsidised by the DFG (INST 211/667-1).

\printbibliography

\end{document}